\documentclass[fleqn,twoside]{article}
\usepackage{espcrc2}
\usepackage{epsfig}
\usepackage{latexsym}
\usepackage{amssymb}

\title{{The finite temperature transition for 3-flavour lattice QCD at finite
isospin density}\thanks{Talk presented by D.~K.~Sinclair, Lattice2004, 
Fermilab.}}

\author{J.~B.~Kogut\address{Department of Physics, University of Illinois,
1110 West Green Street, Urbana, IL 61801, USA}\thanks{Supported in part by NSF
grant NSF PHY03-04252.} and
D.~K.~Sinclair, \address{HEP Division, Argonne National Laboratory, 9700 South
Cass Avenue, Argonne, IL 60439, USA}\thanks{Supported by DOE contract
W-31-109-ENG-38.}}

\begin{document}

\begin{abstract}
We simulate 3-flavour lattice QCD with a small chemical potential $\mu_I$
for isospin, at temperatures close to the finite temperature transition. Using
quark masses just above the critical mass for zero chemical potential, we
determine the position of the transition from hadronic matter to a quark-gluon 
plasma as a function of $\mu_I$. We see evidence for a critical endpoint where
the transition changes from a crossover to a first-order transition as $\mu_I$
is increased. We argue that QCD at finite $\mu_I$ and QCD at finite quark-number
chemical potential $\mu$ should behave similarly in this region.
\end{abstract}

\maketitle

\section{Introduction}

QCD at finite temperature and quark-number chemical potential, $\mu$, is of
relevance to the physics of relativistic heavy-ion collisions. Different
parts of the phase diagram are accessible to the various heavy-ion accelerators
--- RHIC, CERN heavy-ion, AGS and SIS.

For the region of small $\mu$ and temperatures close to the finite temperature
transition, various techniques have been developed to circumvent the problems
associated with the complex fermion determinant at finite $\mu$
\cite{Fodor:2001au,Allton:2002zi,deForcrand:2002ci,D'Elia:2002gd}. We have 
chosen to study the finite temperature transition with a finite chemical
potential $\mu_I$ for isospin, which is equivalent to simulating QCD at finite
$\mu$ with $2\mu=\mu_I$ using only the magnitude of the determinant and
ignoring the phase. Because the fermion phase is relatively well behaved at
small $\mu$, it can be argued that the positions of the transitions and
probably their nature with and without this phase, should be the same.

For 2 flavours we found no sign of a critical endpoint for $\mu_I < m_\pi$,
the maximal range over which finite $\mu_I$ and finite $\mu$ transitions could
be expected to be related \cite{Kogut:2004zg}. 
Here we describe our simulations for 3 flavours.
Not only is this more physical, but for $m < m_c(0)$, the finite temperature 
transition is first order at $\mu_I=0$. At $m = m_c(0)$ it becomes second-order,
in the Ising universality class. For $m > m_c(0)$ the $\mu_I=0$ transition is
a crossover. It is expected that the critical endpoint moves continuously from
$\mu_I=0$ for $m = m_c(0)$ to finite $\mu_I$, as $m$ is increased. Thus we
should be able to keep this endpoint as close to $\mu_I=0$ as we please, by
choosing $m$ to lie just above $m_c(0)$. The preliminary results of our
simulations at such a mass, reported here, do show such a critical endpoint at
a modest value of $\mu_I$.

In section~2 we describe our simulations and report preliminary results. Our
conclusions are given in section~3. 

\section{3-flavour QCD at finite $\mu_I$ and $T$}

The staggered fermion part of our lattice action is
\begin{equation}
S_f=\sum_{sites} \left[\bar{\chi}[D\!\!\!\!/(\frac{1}{2}\tau_3\mu_I)+m]\chi
                   + i\lambda\epsilon\bar{\chi}\tau_2\chi\right].
\end{equation}
For these simulations, which are at $\mu_I < m_\pi$, we set the symmetry 
breaking parameter $\lambda=0$.

We are performing simulations on $8^3 \times 4$ and $12^3 \times 4$ lattices
at $m=0.03$, $0.035$ and $0.04$. [For $N_t=4$, $m_c(0)=0.0331(12)$
\cite{Karsch:2000kv}.] We use
standard hybrid molecular-dynamics simulations. Since with such small lattices
one can get 2-state signals even for crossovers, we use fourth-order Binder
cumulants ($B_4$) \cite{binder}
for $\bar{\psi}\psi$ to get quantitative information about the transition:
\begin{equation}
B_4(X) = {\langle ( X - \langle X \rangle)^4 \rangle \over
          \langle ( X - \langle X \rangle)^2 \rangle^2},
\end{equation}
where X is any observable. Since we use 5 noise vectors for each configuration
to estimate $\bar{\psi}\psi$ (and the number densities), we are able to obtain
an unbiased estimate of $B_4(\bar{\psi}\psi)$, as well as the susceptibility
$\chi_{\bar{\psi}\psi}$. 
\begin{equation}
\chi_{_{\scriptstyle X}} = V \langle X^2 - \langle X \rangle^2 \rangle.
\end{equation}
The $\beta$ of the transition is determined by the positions of the maxima of
the various susceptibilities, or from the minimum of $B_4(\bar{\psi}\psi)$. (We
have checked the consistency of these two methods.)

\begin{figure}[htb]
\vspace{-0.1in}
\epsfxsize=3in
\centerline{\epsffile{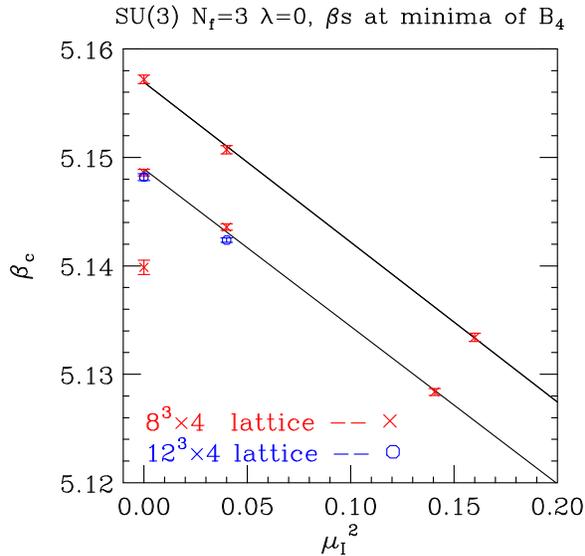}}
\vspace{-0.1in}
\caption{$\beta_c$ as functions of $\mu_I^2$ from minima of $B_4$. The bottom
point is for $m=0.03$. The middle set of points and line are for $m=0.035$.
The top set of points and line are for $m=0.04$.}
\label{fig:beta}
\end{figure}
Figure~\ref{fig:beta} shows the transition $\beta$s, $\beta_c$, as functions
of $\mu_I$ for the $3$ quark masses. The fits shown are
\begin{eqnarray}
m=0.035 &:&\; \beta_c = 5.1489(5) - 0.145(6) \mu_I^2 \nonumber \\
m=0.040 &:&\; \beta_c = 5.1569(3) - 0.147(4) \mu_I^2 
\end{eqnarray}
which reflect the expected analyticity in $\mu_I^2$. $\beta_c$ was obtained
in each case from the position of the minimum of $B_4(\bar{\psi}\psi)$.
Ferrenberg-Swendsen reweighting was used to interpolate between the $\beta$s
of our simulations. These positions were in good agreement with those obtained
from the maxima of the corresponding susceptibilities.

\begin{figure}[htb]
\vspace{-0.1in}
\epsfxsize=3in
\centerline{\epsffile{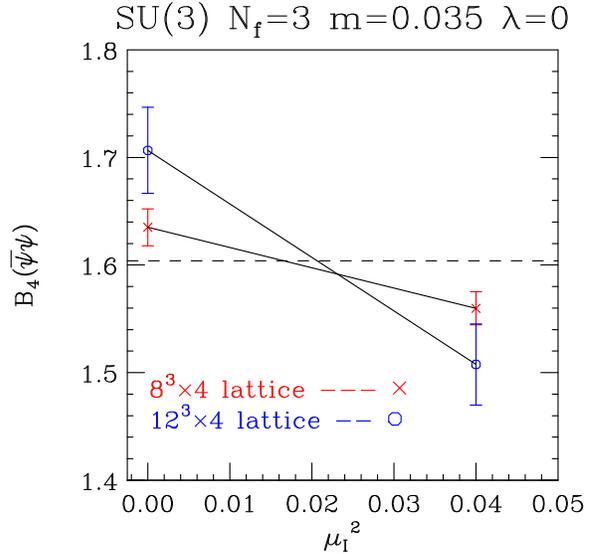}}
\vspace{-0.1in}
\caption{Binder cumulants for $m=0.035$ as functions of $\mu_I$. The dashed
line is at $B_4=1.604$, the 3-d Ising value.}
\label{fig:binder0.035}
\end{figure}
For a lattice of infinite volume, the Binder cumulant $B_4(\bar{\psi}\psi)$
is $3$ for a crossover and $1$ for a first-order transition. The critical
endpoint is expected to be in the universality class of the 3-d Ising model,
for which $B_4(\bar{\psi}\psi)=1.604(1)$. While $B_4$ for a crossover or a
first-order transition approaches its infinite volume limit relatively slowly
with volume, the approach for a second-order phase transition is fast and
approaches the limiting value on relatively small lattices 
\cite{Karsch:2001nf}. Figure~\ref{fig:binder0.035} shows the Binder cumulant 
at the transition for our runs at $m=0.035$ with $\mu_I=0$ and $\mu_I=0.2$ for 
both $8^3 \times 4$ and $12^3 \times 4$ lattices. Despite the sizable error 
bars, these are the results of runs of 160,000 trajectories at the $\beta$ 
closest to the transition.

$B_4$ at $\mu_I=0$ lies above the Ising value while that at $\mu_I=0.2$ lies 
below this value for both lattice sizes. Straight lines through these 2 points
for the two lattice sizes cross at some $\mu_I$ with $0.0 < \mu_I < 0.2$, and
at a $B_4$ value close to the Ising value. These preliminary results 
would indicate that there is a critical endpoint at $\mu_I = \mu_{IE}$ with 
$0.0165 \lesssim \mu_{IE}^2 \lesssim 0.0231$.

For the $m$, $\mu_I$ values at which we have run and for $\beta$s in the
neighbourhood of the transition on an $8^3 \times 4$ lattice we estimate that
the phase of the fermion determinant $\theta$ obeys,
\begin{equation}
\langle \theta^2 \rangle \approx (5\,\mbox{--} 12) \mu_I^2.
\end{equation}
We have argued in \cite{Kogut:2004zg} that 
$\langle \cos \theta \rangle > 0.5$ is a reasonable
estimate of the range over which one can neglect the phase of the fermion
determinant in determining the position and nature of the finite temperature
transition \cite{Kogut:2004zg}. 
If so, the finite $\mu_I$ and $\mu$ behaviour of this 
transition should agree for $\mu_I \le 0.25-0.45$, which includes our estimate
for the critical endpoint for $m=0.035$.

Using $m_c(0)=0.0331$ we then predict that
\begin{equation}
{m(\mu_I) \over m(0)} \approx 1 + 7 \left({\mu_I \over 2 \pi T_c}\right)^2
\end{equation}

\section{Conclusions}

We are simulating 3-flavour lattice QCD at finite $\mu_I$ and temperatures 
close to the finite temperature transition.

We observe the slow decrease of $\beta$ and hence temperature, of the finite
temperature transition with $\mu_I$. Arguing that, in this range of $\mu_I$,
the dependence on $\mu_I$ and $2\mu$ should be the same we find a falloff about
20\% slower than that observed by de Forcrand and Philipsen 
\cite{deForcrand:2002ci} At least some of
the difference is due to omitting ${\cal O}(\mu_I^4)$ terms.

We have obtained preliminary evidence for a critical endpoint for $m=0.035$
(just above $m_c(0)$) from simulations at $\mu_I \leq 0.2$. However,
simulations on an $8^3 \times 4$ lattice at $\mu_I=0.375$ yield $B_4=1.64(2)$,
which does not support this interpretation. This is probably because $0.375$ is
too close to $\mu_c=m_\pi$, but this needs to be clarified. We have seen no
evidence for a critical endpoint at $m=0.04$. Whether this is consistent with
what we see at $m=0.03$ should also be checked.

Our prediction for the $\mu_I$ dependence of $m_c(\mu_I)$ appears inconsistent
with both de Forcrand and Philipsen \cite{deForcrand:2002ci} 
and the Bielefeld-Swansea collaboration \cite{Allton:2002zi},
lying in between their inconsistent results.

Clearly much more needs to be done.

\section*{Acknowledgements}
DKS would like to thank Philippe de Forcrand for helpful discussions on Binder
cumulants. These simulations were performed on the LCRC Jazz cluster at
Argonne, the Tungsten cluster at NCSA and local PCs.

\end{document}